%% file: trustednoise_arxiv_v3.tex
\documentclass[10pt,a4paper,fleqn]{scrartcl}
\pdfoutput=1
\usepackage[utf8]{inputenc}
\usepackage[english]{babel}
\usepackage{amsmath}
\usepackage{amsfonts}
\usepackage{amssymb}
\usepackage{braket}
\usepackage[]{graphicx}
\usepackage{bbold}
\usepackage{appendix}
\usepackage{siunitx}
\usepackage{mathtools}
\usepackage{pbox}
\usepackage{hyperref}
\usepackage{authblk}
\usepackage{subcaption}
\usepackage{url}
\usepackage{cite}

\usepackage{color}

\DeclareMathOperator{\Tr}{Tr}
\DeclareMathOperator{\diag}{diag}
\usepackage[left=1.7cm,right=1.7cm,top=2cm,bottom=2cm]{geometry}

\title{Analysis of the trusted-device scenario in continuous-variable quantum key distribution}
\date{}

\author[*]{\large Fabian Laudenbach}
\author[ ]{\large Christoph Pacher}

\affil[ ]{\normalsize Security \& Communication Technologies, Center for Digital Safety \& Security,\protect \\AIT Austrian Institute of Technology GmbH, Giefinggasse 4, 1210 Vienna, Austria \protect \\ \quad}

\affil[*]{\normalsize mail: fabian.laudenbach@ait.ac.at}

\begin{document}

\maketitle

\input{maintext.tex}

\end{document}

%% file: maintext.tex
\begin{abstract}
\textbf{Abstract.} The assumption that detection and/or state-preparation devices used for continuous-variable quantum key distribution (CV-QKD) are beyond influence of potential eavesdroppers leads to a significant performance enhancement in terms of achievable key rate and transmission distance. We provide a detailed and comprehensible derivation of the Holevo bound in this so-called trusted-device scenario. Modelling an entangling-cloner attack and using some basic algebraic matrix transformations, we show that the computation of the Holevo bound can be reduced to the solution of a quadratic equation. As an advantage of our derivation, the mathematical complexity of our solution does not increase with the number of trusted-noise sources. Finally, we provide a numerical evaluation of our results, illustrating the counter-intuitive fact that an appropriate amount of trusted receiver loss and noise can even be beneficial for the key rate.
\end{abstract}

\section{Introduction}

In contrast to quantum key distribution with discrete variables (DV-QKD; encoding of polarisation, phase or time-bin of single photons), continuous-variable QKD exploits the encoding of the position and momentum quadrature of optical Gaussian states. These can be displaced thermal states~\cite{laudenbach2017continuous, grosshans2002continuos, grosshans2003quantum, grosshans2003virtual, weedbrook2004quantum}, squeezed~\cite{cerf2001quantum} or entangled states~\cite{madsen2012continuos}. In particular, since coherent states can be both reliably produced at high rates and efficiently measured using standard coherent detection, CV-QKD with displaced thermal states (i.e.\ noisy coherent states) is valued for its modest technological requirements. Our recent review article~\cite{laudenbach2017continuous} provides a detailed introduction into the mathematical tools and methods required for the security analysis, noise-modelling and parameter estimation in CV-QKD. The so-called trusted-device scenario, however, is only mentioned briefly and incompletely in the above mentioned review (Sec.~7.3). The present manuscript is a supplement to~\cite{laudenbach2017continuous} which compensates for this shortcoming.

We start with some preliminaries and definitions: Although in practical implementations of CV-QKD, the transmitter (Alice, $A$) will prepare randomly distributed coherent states with a modulation variance $V_{\text{mod}}$ using electro-optic modulation, the standard security analysis assumes that the transmitter and the receiver (Bob, $B$) share a sequence of two-mode-squeezed vacuum states (or EPR states) with variance $V=V_{\text{mod}}+1$ (in shot-noise units, SNU). Alice who generates these EPR states in her lab keeps one mode to herself to perform a heterodyne measurement on and transmits the other one to Bob through an insecure quantum channel. We further assume that an eavesdropper's (Eve, $E$) attack on the quantum channel causes a disturbance in Bob's mode in the shape of an additional quadrature variance, labelled as channel excess noise $\xi_{\text{ch}}$, and an attenuation, labelled as channel transmittance $T_{\text{ch}}$. Bob, using imperfect measurement devices, will experience further noise and attenuation which we label as $\xi_{\text{rec}}$ and $T_{\text{rec}}$. The receiver noise $\xi_{\text{rec}}$ might be composed of the electronic noise of the homodyne detector(s), quantisation errors caused by analogue-to-digital conversion of the measurement outcomes, phase- and intensity noise of the local oscillator and others. The receiver transmission $T_{\text{rec}}$ is the product of the detection and coupling efficiency in Bob's lab. In the trusted-receiver scenario, $\xi_{\text{rec}}$ and $T_{\text{rec}}$ are well-known and calibrated and assumed to be beyond influence of potential eavesdroppers.

Contrary to the more common notation in CV-QKD literature where the channel noise refers to the channel input, we define it referring to the channel output, as received by Bob: $\xi_{\text{ch}}:=T_{\text{ch}}  \xi_{\text{ch},A}= \xi_{\text{ch},B}/ T_{\text{rec}}$, where $\xi_{\text{ch},A}$ is the channel noise referring to Alice and $\xi_{\text{ch},B}$ is the channel noise as measured by Bob. The measured variance of Bob's quadratures is $T_{\text{tot}}(V-1)+1+\xi_{\text{tot}}$ where $T_{\text{tot}}=T_{\text{ch}}T_{\text{rec}}$ and $\xi_{\text{tot}}=T_{\text{rec}}\xi_{\text{ch}}+\xi_{\text{rec}}$ or, in the presence of preparation noise (discussed in Section~\ref{sec_trustedprep}), $\xi_{\text{tot}}=T_{\text{tot}} \xi_{\text{pr}} + T_{\text{rec}}\xi_{\text{ch}} +\xi_{\text{rec}}$.

In continuous-variable quantum key distribution the lower bound on the asymptotic secure-key rate for reverse reconciliation, assuming collective attacks, is described by~\cite{laudenbach2017continuous, devetak2004distillation}

\begin{equation}
K= f_{\text{sym}} (1-\text{FER})(1-\nu)(\beta I_{AB}-\chi_{EB}),
\end{equation}
where $f_{\text{sym}}$ is the symbol rate, $\text{FER}$ is the frame-error rate during error correction, $\nu$ is the fraction of the raw key consumed by parameter estimation, $\beta$ is the reconciliation efficiency, $I_{AB}$ is the mutual information between Alice and Bob and $\chi_{EB}$ is an upper bound for the mutual information between Eve and Bob, also referred to as Holevo bound. In the case of Gaussian modulation of coherent states the mutual information between Alice and Bob reads~\cite{laudenbach2017continuous}

\begin{align}
I_{AB}&=\frac{\mu}{2} \log_{2} \left( 1+\text{SNR} \right) \notag \\
& = \frac{\mu}{2} \log_{2} \left( 1+\frac{T_{\text{tot}}(V-1)}{ \mu +\xi_{\text{tot}}} \right),
\end{align}
where $T_{\text{tot}}$ is the total attenuation (channel and detection), $V$ is the variance of the two-mode-squeezed vacuum state (in SNU) that Alice and Bob share, $\xi_{\text{tot}}$ is the total excess noise as measured by Bob (in SNU) and $\mu$ indicates whether homodyne detection (measurement of $q$ \emph{or} $p$; $\mu=1$) or heterodyne detection (simultaneous measurement of $q$ \emph{and} $p$; $\mu=2$) is performed.

The Holevo bound of a CV-QKD system with reverse reconciliation is the difference between Eve's von Neumann entropy $S$ before and after Bob performed a projective quadrature measurement:
\begin{equation}
\chi_{EB}=S_{E}-S_{E|B}.
\end{equation}
In general, the von Neumann entropy is computed using

\begin{equation} \label{eq_vonneumann}
S=\sum_{i} \left( \frac{\nu_{i}+1}{2} \log_{2} \left( \frac{\nu_{i}+1}{2} \right) - \frac{\nu_{i}-1}{2} \log_{2} \left( \frac{\nu_{i}-1}{2} \right) \right),
\end{equation}
where $\nu_{i}$ are the symplectic eigenvalues of the covariance matrices that describe the information accessible to Eve. For a Gaussian state with $N$ modes the symplectic eigenvalues correspond to the elements of the diagonalised covariance matrix when decomposed into its Williamson form~\cite{weedbrook2012gaussian}:

\begin{equation}
\Sigma=\mathcal{S} \bigoplus_{i=1}^{N}
\begin{pmatrix}
\nu_{i} & 0 \\
0 & \nu_{i} 
\end{pmatrix}
\mathcal{S}^{T}.
\end{equation}
Here $\mathcal{S} \in \mathbb{R}^{2N \times 2N}$ is a symplectic matrix, i.e.\ it fulfils the condition

\begin{equation}
\mathcal{S} \Omega \mathcal{S}^{T}=\Omega,
\end{equation} 
where $\Omega$ is the so-called symplectic form:

\begin{equation} \label{eq_Omega}
\Omega=\bigoplus_{i=1}^{N}
\begin{pmatrix}
0 & 1 \\
-1 & 0
\end{pmatrix}.
\end{equation}
In general, the symplectic eigenvalues of a covariance matrix $\Sigma$ correspond to the positive eigenvalues of $i \Omega \Sigma$.

When modelling Eve's attack it is assumed that she \emph{purifies} the state shared by Alice and Bob:

\begin{equation}
\rho_{AB}=\Tr_{E}\left(\rho_{ABE}\right)
\end{equation}
and the total state $\rho_{ABE}=\ket{\Psi_{ABE}}\bra{\Psi_{ABE}}$ is pure. A Gaussian (multimode) state is pure if and only if its covariance matrix has a symplectic rank $\mathcal{R}$ of zero. The symplectic rank is the number of symplectic eigenvalues different from 1.

It is important to remark that for the calculation of the Holevo bound it does not make a difference how exactly Eve purifies $\rho_{AB}$. This is because for any two states $\rho_{E_{1}}$ and $\rho_{E_{2}}$ that both purify $\rho_{AB}$ there will be a unitary transformation $U$ that can transform one into the other:

\begin{equation}
(\mathbb{1}_{AB} \otimes U_{E})\ket{\Psi_{ABE_{1}}}=\ket{\Psi_{ABE_{2}}}.
\end{equation}
This `freedom in purification'~\cite{nielsen2000quantum} is particularly useful since unitary transformations are entropy-preserving. Thus we have

\begin{equation}
S(\rho_{E_{1}})=S(U_{E} \rho_{E_{1}} U_{E}^{\dagger})=S(\rho_{E_{2}}),
\end{equation}
or, in other words, Eve's entropy (and therefore the Holevo bound) does not depend on the way she purifies Alice's and Bob's mutual state.

In the language of Gaussian quantum information, the freedom in purification can be expressed as follows: The covariance matrix of Alice's and Bob's mutual state has a symplectic rank $\mathcal{R}(\Sigma_{AB})>0$. Eve purifies the state such that $\mathcal{R}(\Sigma_{ABE})=0$ (all symplectic eigenvalues equal to one). Different purifications can always be transformed into each other by a symplectic transformation acting on Eve's subsystem:

\begin{equation}
(\mathbb{1}_{AB} \oplus \mathcal{S}_{E})\Sigma_{ABE_{1}}(\mathbb{1}_{AB} \oplus \mathcal{S}_{E})^{T}=\Sigma_{ABE_{2}},
\end{equation}
leaving the entropy invariant.

\begin{figure}
\centering
\subcaptionbox{}
    [\linewidth]{\includegraphics[width=\linewidth]{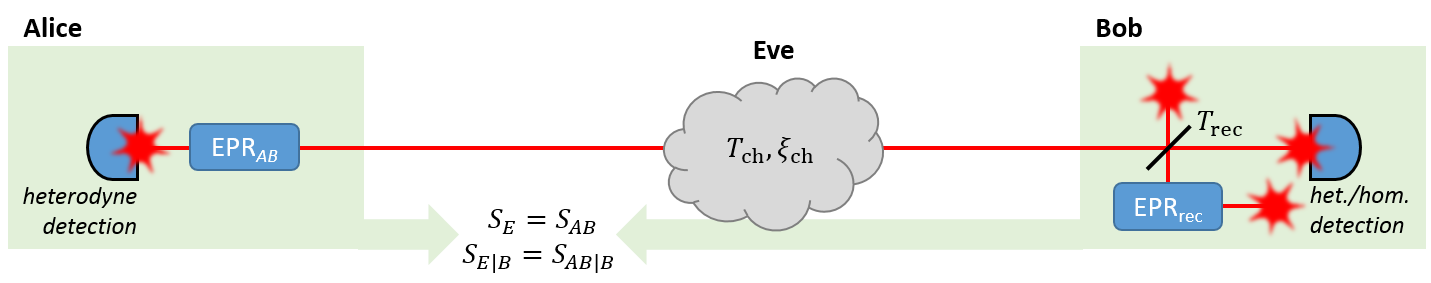}}
\subcaptionbox{}
    [\linewidth]{\includegraphics[width=\linewidth]{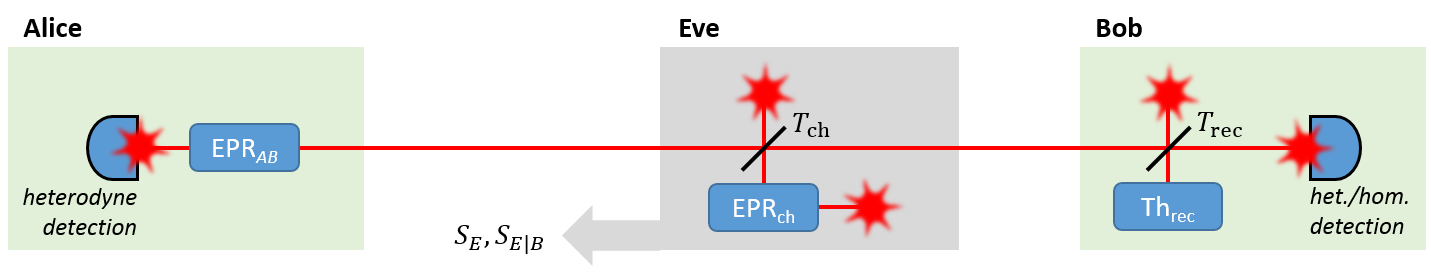}}
\caption{Two different approaches to compute $S_{E}$ and $S_{E|B}$ in the trusted-receiver assumption. The purification ansatz (a) does not make any assumptions on the particular eavesdropping attack. Eve's interaction with the quantum channel will modify the initial quantum state to \eqref{eq_covABloose}. The trusted receiver is modelled by an additional two-mode squeezed vacuum state interacting with Bob's mode. The total state therefore comprises four optical modes (star-shaped red sparks). A measurement of Bob's mode will reduce the total state to three modes, described by a $6\times6$ covariance matrix. Eve is not directly represented in this matrix. Since we assume her to hold a purification of Alice's and Bob's state, her conditional entropy $S_{E|B}$ equals the one of the remaining total state. In the entangling-cloner ansatz (b), the channel itself is modelled with an EPR state controlled by Eve. Receiver loss and noise are now modelled by a thermal state. The total state now comprises five optical modes before and four modes after Bob's projective measurement. As opposed to the purification ansatz, Eve is now represented in the total state by her own state EPR$_{\text{ch}}$. In order to compute her entropy it is sufficient to compute the eigenvalues of her subsystem, represented by a $4\times4$ matrix (as opposed to a $6\times6$ matrix in the purification ansatz). We show below how this problem can be even further reduced to dimension $2\times2$.}
\label{comparison}
\end{figure}

In this paper we discuss two approaches to model the impact of trusted devices on the Holevo bound: The first ansatz (Section~\ref{sec_purif}) is based on the fact that for pure bipartite states the entropy is the same in both subsystems (we will slovenly refer to this method as `purification ansatz'); the second ansatz (Section~\ref{sec_entang}) is based on an entangling-cloner attack. Both approaches are illustrated in Figure~\ref{comparison}. Since in both approaches Eve purifies Alice's and Bob's subsystem, they are both equivalent and lead to the same results due to the freedom in purification, as explained above. Assuming a trusted receiver, we show that the former approach requires to find the eigenvalues of a $6\times6$ matrix whereas in the latter approach the problem is reduced to a $4\times4$ matrix. Using some basic algebraic matrix transformations, we show that the eigenvalue problem of the entangling-cloner attack can be further reduced to a $2\times2$ matrix, allowing for analytic expressions of manageable complexity. Furthermore, as elaborated in Section~\ref{sec_trustedprep}, our approach reduces the mathematical complexity even more drastically in a trusted-receiver \emph{and} -preparation scenario.

The paper is organised as follows: Section~\ref{sec_purif} sketches the purification ansatz to incorporate the trusted receiver into the security analysis. This section does not contain any novel results as it is just a review of a well-established method; it is discussed here merely for the sake of completeness and comparison to our own method. Section~\ref{sec_entang} provides a detailed step-by-step derivation of the symplectic eigenvalues for the case of an entangling-cloner attack. As we point out, this method always leads to quadratic equations regardless of the number of trusted-noise sources. This allows us to derive an analytic solution for the case of trusted state preparation and receiver -- a result that, to the best of our knowledge, has not been published before. In Section~\ref{sec_numerical} we perform some numerical simulations to provide examples and illustrations on how different security assumptions affect the performance of practical CV-QKD systems. Moreover, we discuss how deliberately detuning the receiver's quantum efficiency can be used to match the signal-to-noise ratio to a given error-correction code. Finally, we conclude in Section~\ref{sec_concl}.

\section{Purification ansatz} \label{sec_purif}

After purification of $\rho_{AB}$ the total state can be seen as a pure bipartite state with Alice's and Bob's mutual subsystem on one side and Eve's subsystem on the other. As such it can be written as a Schmidt decomposition:

\begin{equation}
\ket{\Psi_{ABE}}=\sum_{j} \lambda_{j} \ket{\psi_i}_{AB}\ket{\phi_i}_{E}.
\end{equation}
Sharing the coefficients $\lambda$, both subsystems have the same entropy. Thus we have $S_{E}=S_{AB}$ and $S_{AB}$ is obtained from the symplectic eigenvalues of a covariance matrix of the form (e.g.~derived in~\cite[Appendix~C]{laudenbach2017continuous})

\begin{equation} \label{eq_covABstrict}
\Sigma_{AB}=
\begin{pmatrix}
V \mathbb{1}_{2} & \sqrt{T(V^{2}-1)} \sigma_{z} \\ 
\sqrt{T(V^{2}-1)} \sigma_{z} & (T(V-1)+1+\xi) \mathbb{1}_{2}
\end{pmatrix}.
\end{equation}
If the entire transmission $T$ and excess noise $\xi$ are attributed to Eve, the parameters above comprise both the contributions from the channel as well as from the receiver, i.e.\ $T=T_{\text{ch}}T_{\text{rec}}$ and $\xi=T_{\text{rec}}\xi_{\text{ch}}+\xi_{\text{rec}}$, where $\xi_{\text{ch}}$ is the channel noise as received by Bob. 

Under the relaxed assumption that the detection devices in Bob's lab are well calibrated and trusted, $T_{\text{rec}}$ and $\xi_{\text{rec}}$ are beyond Eve's influence and the covariance matrix describing her von Neumann entropy before Bob's measurement reads

\begin{equation} \label{eq_covABloose}
\Sigma_{AB}^{\text{trusted rec.}}=
\begin{pmatrix}
V \mathbb{1}_{2} & \sqrt{T_{\text{ch}}(V^{2}-1)} \sigma_{z} \\ 
\sqrt{T_{\text{ch}}(V^{2}-1)} \sigma_{z} & (T_{\text{ch}}(V-1)+1+\xi_{\text{ch}}) \mathbb{1}_{2}
\end{pmatrix}.
\end{equation}
Depending on whether the receiver is assumed to be trusted or not, Eves entropy $S_{E}$ will be computed using the matrix~\eqref{eq_covABloose} or \eqref{eq_covABstrict}. Since both matrices are of the form

\begin{equation} \label{eq_covform}
\begin{pmatrix}
 a \mathbb{1}_{2} & c \sigma_{z} \\
 c \sigma_{z} & b \mathbb{1}_{2} \\
\end{pmatrix} .
\end{equation}
their symplectic eigenvalues can be obtained by~\cite{weedbrook2012gaussian}

\begin{equation} \label{eq_nu12}
\nu_{1,2}=\frac{1}{2} \left(z \pm [b-a] \right) \qquad \text{with} \quad z=\sqrt{(a+b)^{2}-4c^{2}} .
\end{equation}
Inserting $\nu_{1}$ and $\nu_{2}$ into~\eqref{eq_vonneumann} yields Eve's von Neumann entropy $S_{E}$.

Although in the trusted-receiver model $T_{\text{rec}}$ and $\xi_{\text{rec}}$ do not contribute to $S_{E}$, they still influence Bob's measurement and therefore also Eve's entropy conditioned on Bob's measurement. The computation of $S_{E|B}$ is therefore a bit more elaborate than just omitting $T_{\text{rec}}$ and $\xi_{\text{rec}}$ in the calculations (and has not been discussed in our review paper~\cite{laudenbach2017continuous}).

The purification-based approach to compute $S_{E|B}$ with a trusted receiver, which is in more detail described in \cite{lodewyck2007quantum, fossier2009improvement, usenko2016trusted}, assumes an unspecified eavesdropper attack on the quantum channel leading to the covariance matrix~\eqref{eq_covABloose}. This attack assumes Eve to purify the \emph{total} state. However, since in the trusted-device scenario she can only purify the channel, the ancillary state that we use to introduce imperfect detection (to which Eve has no access to) needs to be a pure state itself. Therefore the detector is modelled by an additional EPR state (as EPR states purify thermal states), maintaining the purity of the total state. The imperfect detection is modelled by mixing Bob's mode of the shared EPR state~\eqref{eq_covABloose} with one mode of the ancillary EPR state using a beamsplitter of transmittance $T_{\text{rec}}$ in Bob's lab, as illustrated in Figure~\ref{comparison}(a). The total state therefore consists of four optical modes and is represented by an $8 \times 8$ matrix. A projective quadrature measurement of Bob's mode will reduce the total state to three modes, i.e.\ a $6 \times 6$ covariance matrix labelled as $\Sigma_{\text{tot}|B}$. Depending on whether Bob performs heterodyne or homodyne detection, this state is given by~\cite{laudenbach2017continuous}

\begin{subequations}
\begin{align}
\Sigma_{\text{tot}|B} (\text{het}) &  = \Sigma_{A,\text{rec}}  - \frac{1}{V_{B}+1} \Sigma_{C} \Sigma_{C}^{T} , \label{eq_partialhet} \\  
\Sigma_{\text{tot}|B} (\text{hom}) & = \Sigma_{A,\text{rec}} - \frac{1}{V_{B}} \Sigma_{C} \Pi_{q,p}  \Sigma_{C}^{T} ,  \label{eq_partialhom}
\end{align}
\end{subequations}
where $\Sigma_{A,\text{rec}} \in \mathbb{R}^{6 \times 6}$ describes the submatrix representing Alice's half of the initial EPR pair and the EPR pair used to model the imperfect receiver, $\Sigma_{C} \in \mathbb{R}^{6 \times 2}$ is the submatrix describing the quadrature correlations of $\Sigma_{A,\text{rec}}$ with Bob's mode, $V_{B}$ is Bob's quadrature variance and $\Pi$ is a projection operator defined as

\begin{equation}
\Pi_{q}=
\left(
\begin{array}{cc}
 1 & 0 \\
 0 & 0 \\
\end{array}
\right) \quad \text{$q$-measurement}, \\
\Pi_{p}=
\left(
\begin{array}{cc}
 0 & 0 \\
 0 & 1 \\
\end{array}
\right) \quad \text{$p$-measurement}.
\end{equation}
In the projected state $\Sigma_{\text{tot}|B}$ there are no optical modes attributed to Eve. The three remaining modes are inaccessible to her. However, assuming that she purifies the total state, it is sufficient to know the entropy of $\Sigma_{\text{tot}|B}$ in order to obtain $S_{E|B}$ since it coincides with Eve's entropy. This is analogous to the computation of $S_{E}$ where knowledge of the matrix~\eqref{eq_covABstrict} (untrusted receiver) or \eqref{eq_covABloose} (trusted receiver) allows for the derivation of Eve's entropy before the projective measurement.

The matrix $\Sigma_{\text{tot}|B}$ is of dimension $6 \times 6$ but can be rearranged into a block-diagonal representation of two equivalent $3 \times 3$ matrices using similarity transformations analogous to the ones described in the subsequent section. The eigenvalue problem can thereby be reduced to a cubic equation.

\section{Entangling-cloner ansatz} \label{sec_entang}

This ansatz has been previously described in \cite{laudenbach2017continuous, weedbrook2012continuous}, however only for the scenario of an untrusted receiver. In this approach the trusted-receiver scenario can be modelled as follows: The total state now consists of two EPR states and a thermal state (see Figure~\ref{comparison}(b)) which are each uniquely defined by their variance: one entangled state $\text{EPR}_{AB}$ with variance $V$ used for key exchange by Alice and Bob, one entangled state $\text{EPR}_{\text{ch}}$ with variance $W_{\text{ch}}$ modelling noise and loss in the quantum channel and a thermal state $\text{Th}_{\text{rec}}$ with variance $W_{\text{rec}}$ modelling receiver noise and loss. Beamsplitters, one with transmission $T_{\text{ch}}$ and one with transmission $T_{\text{rec}}$ mix Bob's mode of the initial EPR state with a channel mode and the thermal state, respectively. The total state before action of the beamsplitters can be represented by the covariance matrix

\begin{equation}
\Sigma_{\text{tot},0}=
\left(
\begin{array}{cccccc}
 V  \mathbb{1}_{2} & \sqrt{V^2-1} \sigma_{z} & 0 & 0 & 0 \\
 \sqrt{V^2-1} \sigma_{z} &  V \mathbb{1}_{2} & 0 & 0 & 0 \\
 0 & 0 & W_{\text{ch}} \mathbb{1}_{2} &\sqrt{W_{\text{ch}}^2-1} \sigma_{z} & 0 \\
 0 & 0 & \sqrt{W_{\text{ch}}^2-1} \sigma_{z} & W_{\text{ch}} \mathbb{1}_{2} & 0 \\
 0 & 0 & 0 & 0 & W_{\text{rec}} \mathbb{1}_{2} \\
\end{array}
\right),
\end{equation}
which can be more handily written in terms of the the direct sum:

\begin{equation}
\Sigma_{\text{tot},0}= \text{EPR}_{AB} \oplus \text{EPR}_{\text{ch}} \oplus \text{Th}_{\text{rec}}.
\end{equation}
We remark at this point that Eve's attack on the quantum channel involves only the two noise- and lossless entangled states $\text{EPR}_{AB}$ and $\text{EPR}_{\text{ch}}$ and is therefore pure. Freedom in purification (as argued in the Introduction) guarantees the equivalence to any other purification-based attack in terms of Eve's entropy. In contrast to Sec.~\ref{sec_purif}, it is therefore sufficient to model the noisy receiver with a single-mode thermal state.

A beamsplitter located at the channel acts on Bob's mode and the first mode of the channel EPR state; a second beamsplitter located at the receiver acts on Bob's mode and the thermal state modelling the receiver:

\begin{subequations}
\begin{align}
\text{BS}_{\text{ch}} &=
\left(
\begin{array}{cccccc}
 \mathbb{1}_{2} & 0 & 0 & 0 & 0 \\
 0 & \sqrt{T_{\text{ch}}} \mathbb{1}_{2} & \sqrt{1-T_{\text{ch}}} \mathbb{1}_{2} & 0 & 0 \\
 0 & - \sqrt{1-T_{\text{ch}}} \mathbb{1}_{2} & \sqrt{T_{\text{ch}}} \mathbb{1}_{2} & 0 & 0 \\
 0 & 0 & 0 & \mathbb{1}_{2} & 0 \\
 0 & 0 & 0 & 0 & \mathbb{1}_{2} \\
\end{array}
\right), \\
\text{BS}_{\text{rec}} &=
\left(
\begin{array}{cccccc}
 \mathbb{1}_{2} & 0 & 0 & 0 & 0 \\
 0 & \sqrt{T_{\text{rec}}} \mathbb{1}_{2} & 0 & 0 & \sqrt{1-T_{\text{rec}}} \mathbb{1}_{2} \\
 0 & 0 & \mathbb{1}_{2} & 0 & 0 \\
 0 & 0 & 0 & \mathbb{1}_{2} & 0 \\
 0 & - \sqrt{1-T_{\text{rec}}} \mathbb{1}_{2} & 0 & 0 & \sqrt{T_{\text{rec}}} \mathbb{1}_{2} \\
\end{array}
\right).
\end{align}
\end{subequations}
Labelling the subsequent action of both beamsplitters as $\text{BS}_{\text{tot}}=\text{BS}_{\text{rec}}\text{BS}_{\text{ch}}$, the total quantum state transforms as follows:

\begin{align}
\Sigma_{\text{tot}}=\text{BS}_{\text{tot}} \ \Sigma_{\text{tot},0} \ \text{BS}_{\text{tot}}^{T}.
\end{align}
We omit writing down the resulting $10 \times 10$ matrix which is rather bulky and little illuminating. Instead, we first concentrate on the subsystem shared by Alice and Bob, i.e.\ the first two rows and columns of $\Sigma_{\text{tot}}$ (four if the block matrices $\mathbb{1}_{2}$ and $\sigma_{z}$ are expanded):

\begin{equation} \label{eq_covAB1}
\Sigma_{AB}=
\left(
\begin{array}{cc}
 V\mathbb{1}_{2} &  \sqrt{T_{\text{ch}}} \sqrt{T_{\text{rec}}} \sqrt{V^2-1} \sigma_{z} \\
 \sqrt{T_{\text{ch}}} \sqrt{T_{\text{rec}}} \sqrt{V^2-1} \sigma_{z} & (T_{\text{ch}} T_{\text{rec}}V +(1-T_{\text{ch}}) T_{\text{rec}} W_{\text{ch}} +(1-T_{\text{rec}}) W_{\text{rec}}) \mathbb{1}_{2} \\
\end{array}
\right)
\end{equation}
Now, defining the variances of the EPR states such that

\begin{subequations}
\begin{align}
W_{\text{ch}} & =\frac{\xi_{\text{ch}}}{1-T_{\text{ch}}}+1, \\
W_{\text{rec}} & =\frac{\xi _{\text{rec}}}{1-T_{\text{rec}}}+1,
\end{align}
\end{subequations}
the variance of Bob's mode becomes

\begin{align} \label{eq_VB}
V_{B}(q)=V_{B}(p)&=T_{\text{ch}} T_{\text{rec}}(V-1)+1+T_{\text{rec}}\xi _{\text{ch}}+\xi _{\text{rec}} \notag \\
& = T_{\text{ch}} T_{\text{rec}}(V-1)+1+\xi _{\text{ch},B}+\xi _{\text{rec}}
\end{align}
and the covariance matrix~\eqref{eq_covAB1} reads

\begin{equation}
\Sigma_{AB}=
\left(
\begin{array}{cc}
 V\mathbb{1}_{2} &  \sqrt{T_{\text{tot}}} \sqrt{V^2-1} \sigma_{z} \\
 \sqrt{T_{\text{tot}}} \sqrt{V^2-1} \sigma_{z} & (T_{\text{tot}}(V-1)+1+\xi _{\text{tot}}) \mathbb{1}_{2} \\
\end{array}
\right),
\end{equation}
which is consistent with \eqref{eq_covABstrict}.

On the other hand, extracting from $\Sigma_{\text{tot}}$ the two modes belonging to Eve, we obtain

\begin{equation} \label{eq_SE}
\Sigma_{E}=
\left(
\begin{array}{cc}
 ((1-T_{\text{ch}}) V + T_{\text{ch}}W_{\text{ch}}) \mathbb{1}_{2} &  \sqrt{T_{\text{ch}}} \sqrt{W_{\text{ch}}^2-1} \sigma_{z} \\
 \sqrt{T_{\text{ch}}} \sqrt{W_{\text{ch}}^2-1} \sigma_{z} & W_{\text{ch}} \mathbb{1}_{2} \\
\end{array}
\right).
\end{equation}
This matrix is of the shape~\eqref{eq_covform}, and therefore its symplectic eigenvalues can be computed by~\eqref{eq_nu12}. As can be verified, its entropy $S_{E}$ coincides with the one shared by Alice and Bob, obtained from their mutual covariance matrix under loose assumptions~\eqref{eq_covABloose} which is exactly what we expect to see when Eve holds a purification of Alice's and Bob's state:

\begin{equation}
S_{E} := S(\Sigma_{E}) = S(\Sigma_{AB}^{\text{trusted rec.}}) =: S_{AB}.
\end{equation}
Now, in order to obtain $S_{E|B}$ we first need to compute $\Sigma_{\text{tot}|B}$, hence the total state of the remaining modes after a projective measurement of Bob's mode. We first rearrange $\Sigma_{\text{tot}}$ such that Bob's mode is represented in the last row and column. This can be done using the permutation matrix

\begin{equation}
P_{3,4\rightarrow 9,10}=
\left(
\begin{array}{cccccccccccc}
 \mathbb{1}_{2} & 0 & 0 & 0 & 0 \\
 0 & 0 & \mathbb{1}_{2} & 0 & 0 \\
 0 & 0 & 0 & \mathbb{1}_{2} & 0 \\
 0 & 0 & 0 & 0 & \mathbb{1}_{2} \\
 0 & \mathbb{1}_{2} & 0 & 0 & 0 \\
\end{array}
\right)
\end{equation}
which will permute the third and fourth row (column) to the bottom (right) when multiplied with $\Sigma_{\text{tot}}$ from the front (back):

\begin{equation}
\Sigma_{\text{tot}}'=P_{3,4\rightarrow 9,10}\ \Sigma_{\text{tot}} \ P_{3,4\rightarrow 9,10}^{T} .
\end{equation}
Since $P_{3,4\rightarrow 9,10}P_{3,4\rightarrow 9,10}^{T}=\mathbb{1}$, the above permutation is a similarity transformation and therefore leaves the eigenvalues of $\Sigma_{\text{tot}}$ invariant. The rearranged matrix is now of the form

\begin{equation}
\Sigma_{\text{tot}}'=
\left(
\begin{array}{cc}
 \Sigma_{A,\text{ch,rec}} & \Sigma_{C}  \\
 \Sigma_{C}^{T} & \Sigma_{B}  \\
\end{array}
\right),
\end{equation}
where $\Sigma_{A,\text{ch,rec}} \in \mathbb{R}^{8 \times 8}$ describes Alice's mode and the two EPR states modelling the channel and the receiver, $\Sigma_{B} \in \mathbb{R}^{2 \times 2}$ is Bob's mode and $\Sigma_{C} \in \mathbb{R}^{8 \times 2}$ describes the mutual quadrature correlations between $\Sigma_{A,\text{ch,rec}}$ and $\Sigma_{B}$. The partial matrix after a projective measurement of Bob's mode depends on whether Bob performs heterodyne or homodyne detection.

\subsection{Heterodyne detection}

In the case of heterodyne detection of Bob's mode, the remaining modes are projected into the state described by the $8 \times 8$ matrix

\begin{equation}
\Sigma_{\text{tot}|B}  = \Sigma_{A,\text{ch,rec}}  - \frac{1}{V_{B}+1} \Sigma_{C} \Sigma_{C}^{T} .
\end{equation}
Again, it is not necessary to evaluate the entire resulting matrix. Instead we only calculate the block matrix that describes Eve's information, i.e.\ the two modes representing the EPR state that was used to model the channel noise and transmission. This matrix reads

\begin{equation} \label{eq_E|Bhet}
\Sigma_{E|B}=
\frac{1}{V_{B}+1}
\left(
\begin{array}{cc}
 e_{1} \mathbb{1}_{2} & e_{2} \sigma_{z}  \\
 e_{2} \sigma_{z} & e_{3} \mathbb{1}_{2} \\
\end{array}
\right)
\end{equation}
with

\begin{subequations} \label{eq_erechet}
\begin{align}
e_{1} & = V \left((1-T_{\text{rec}}) W_{\text{rec}}+T_{\text{rec}} W_{\text{ch}}+1 \right)+T_{\text{ch}} (W_{\text{ch}}-V) \left( 1+(1-T_{\text{rec}}) W_{\text{rec}} \right), \\
e_{2} & = \sqrt{T_{\text{ch}} \left(W_{\text{ch}}^2-1\right)} \left(T_{\text{rec}} V+ (1-T_{\text{rec}})W_{\text{rec}}+1\right), \\
e_{3} & = (1-T_{\text{rec}}) W_{\text{ch}} W_{\text{rec}}+T_{\text{rec}} T_{\text{ch}} (V W_{\text{ch}}-1 ) + T_{\text{rec}} + W_{\text{ch}},
\end{align}
\end{subequations}
and $V_{B}$ is given by \eqref{eq_VB}. Since the above matrix $\Sigma_{E|B}$ is again of the form~\eqref{eq_covform} we obtain the symplectic eigenvalues $\nu_{3}$ and $\nu_{4}$ by using~\eqref{eq_nu12}:

\begin{equation} \label{eq_nu34het}
\nu_{3,4}=\frac{z \pm (e_{3}-e_{1})}{2 (V_{B}+1)}  \qquad \text{with} \quad z=\sqrt{(e_{1}+e_{3})^{2}-4e_{2}^{2}} .
\end{equation}
As can be verified, this result coincides with the one from~\cite{fossier2009improvement} which was obtained using the purification ansatz.

\subsection{Homodyne detection}

In the case of homodyne detection of Bob's mode, the remaining modes are projected into the state

\begin{equation}
\Sigma_{\text{tot}|B} = \Sigma_{A,\text{ch,rec}} - \frac{1}{V_{B}} \Sigma_{C} \Pi_{q,p}  \Sigma_{C}^{T} .
\end{equation}
As in the heterodyne case, we extract from this matrix the two modes controlled by Eve, described by the submatrix $\Sigma_{E|B}$. Depending on whether a $q$- or $p$-measurement has been performed, Eve's state after homodyne measurement of Bob's mode reads:

\begin{equation} \label{eq_E|Bhom}
\Sigma_{E|B}(q)= 
\left(
\begin{array}{cccc}
 e_{1} & 0 & e_{2} & 0 \\
 0 & e_{3} & 0 & e_{4} \\
 e_{2} & 0 & e_{5} & 0 \\
 0 & e_{4} & 0 & e_{6} \\
\end{array}
\right),
\qquad
\Sigma_{E|B}(p)= 
\left(
\begin{array}{cccc}
 e_{3} & 0 & -e_{4} & 0 \\
 0 & e_{1} & 0 & -e_{2} \\
 -e_{4} & 0 & e_{6} & 0 \\
 0 & -e_{2} & 0 & e_{5} \\
\end{array}
\right)
\end{equation}
with

\begin{subequations} \label{eq_erechom}
\begin{align}
e_{1} & = V + \frac{1}{V_{B}}   T_{\text{ch}} (W_{\text{ch}}-V) \left( T_{\text{rec}} V +(1-T_{\text{rec}}) W_{\text{rec}} \right), \\
e_{2} & = \frac{1}{V_{B}}  \sqrt{T_{\text{ch}} (W_{\text{ch}}^2-1)} \left(T_{\text{rec}} V+(1-T_{\text{rec}})W_{\text{rec}} \right) , \\
e_{3} & = V + T_{\text{ch}} \left(W_{\text{ch}}-V\right), \\
e_{4} & = - \sqrt{T_{\text{ch}} \left(W_{\text{ch}}^2-1\right)}, \\
e_{5} & = W_{\text{ch}} - \frac{1}{V_{B}} \left(1-T_{\text{ch}}\right) T_{\text{rec}}  \left(W_{\text{ch}}^2-1\right), \\
e_{6} & =  W_{\text{ch}}.
\end{align}
\end{subequations}
Since a projective homodyne measurement affects the $q$- and $p$-quadratures of the remaining modes differently, $\Sigma_{E|B}$ has now to be described by six independent components, as opposed to only three components in the case of heterodyne detection. Moreover, as this matrix is not similar to~\eqref{eq_covform}, we need to compute the symplectic eigenvalues by hand. We recall that the symplectic eigenvalues of $\Sigma_{E|B}$ are the positive eigenvalues of $i \Omega_{2} \Sigma_{E|B}$ where $\Omega_{2}$ is given by~\eqref{eq_Omega}. This leads to the expression (depending on whether a $q$- or $p$-measurement has been performed)

\begin{align}
\Sigma_{E|B,\text{sympl}}(q)= i
\left(
\begin{array}{cccc}
 0 & e_{3} & 0 & e_{4} \\
 -e_{1} & 0 & -e_{2} & 0 \\
 0 & e_{4} & 0 & e_{6} \\
 -e_{2} & 0 & -e_{5} & 0 \\
\end{array}
\right),
\qquad
\Sigma_{E|B,\text{sympl}}(p)= i
\left(
\begin{array}{cccc}
 0 & e_{1} & 0 & -e_{2} \\
 -e_{3} & 0 & e_{4} & 0 \\
 0 & -e_{2} & 0 & e_{5} \\
 e_{4} & 0 & -e_{6} & 0 \\
\end{array}
\right) .
\end{align}
The matrix $\Sigma_{E|B,\text{sympl}}(p)$ is related to the transpose of $\Sigma_{E|B,\text{sympl}}(q)$ by the similarity transformation

\begin{align}
\Sigma_{E|B,\text{sympl}}(p)=\diag(-1,1,1,-1) \Sigma_{E|B,\text{sympl}}(q)^{T} \diag(-1,1,1,-1)
\end{align}
and therefore has the same eigenvalues. Thus, we can omit the separate consideration of $q$- and $p$-measurement in the following derivation. We now square $\Sigma_{E|B,\text{sympl}}(q)$ (keeping in mind that that the eigenvalues of a squared matrix are just its squared eigenvalues):

\begin{align}
\Sigma^{2}_{E|B,\text{sympl}}(q)&= 
\left(
\begin{array}{cccc}
 e_{1}e_{3}+e_{2}e_{4} & 0 & e_{2}e_{3}+e_{4}e_{5} & 0 \\
 0 & e_{1}e_{3}+e_{2}e_{4} & 0 & e_{1}e_{4}+e_{2}e_{6} \\
 e_{1}e_{4}+e_{2}e_{6} & 0 & e_{2}e_{4}+e_{5}e_{6} & 0 \\
 0 & e_{2}e_{3}+e_{4}e_{5} & 0 & e_{2}e_{4}+e_{5}e_{6} \\
\end{array}
\right).
\end{align}
The advantage of squaring $\Sigma_{E|B,\text{sympl}}(q)$ is that the above matrix can now be brought into a block-diagonal form by rearranging the rows and columns. In particular, swapping the second and third column and row using the permutation matrix

\begin{equation}
P_{2\leftrightarrow 3}=
\left(
\begin{array}{cccc}
 1 & 0 & 0 & 0  \\
 0 & 0 & 1 & 0  \\
 0 & 1 & 0 & 0  \\
 0 & 0 & 0 & 1  \\
\end{array}
\right)
\end{equation}
yields the expression

\begin{align}
P_{2\leftrightarrow 3} \ \Sigma^{2}_{E|B,\text{sympl}}(q) \ P_{2\leftrightarrow 3}&=
\left(
\begin{array}{cccc}
 e_{1}e_{3}+e_{2}e_{4} & e_{2}e_{3}+e_{4}e_{5} & 0 & 0 \\
 e_{1}e_{4}+e_{2}e_{6} & e_{2}e_{4}+e_{5}e_{6} & 0 & 0 \\
 0 & 0 & e_{1}e_{3}+e_{2}e_{4} & e_{1}e_{4}+e_{2}e_{6} \\
 0 & 0 & e_{2}e_{3}+e_{4}e_{5} & e_{2}e_{4}+e_{5}e_{6} \\
\end{array}
\right) =:
\left(
\begin{array}{cccc}
 \mathcal{E} & 0 \\
 0 & \mathcal{E}^{T} \\
\end{array}
\right). 
\end{align}
(Since $P_{2\leftrightarrow 3} P_{2\leftrightarrow 3}=\mathbb{1}$ the operation above is another similarity transformation and therefore does not affect the eigenvalues.) The eigenvalues of a block-diagonal matrix are the union of the eigenvalues of the individual blocks. This particular problem is even more simplified by the observation that the two blocks of each matrix are transposes of each other and therefore have the same eigenvalues. So the whole problem is reduced to finding the eigenvalues of one $2\times2$ matrix. The square roots of these eigenvalues represent the eigenvalues of $\Sigma_{E|B,\text{sympl}}$ which always occur in pairs $\pm\nu_i$. The symplectic eigenvalues of $\Sigma_{E|B}$ that we need for computation of $S_{E|B}$ are the positive eigenvalues of $\Sigma_{E|B,\text{sympl}}$ and given by

\begin{align} \label{eq_nu34hom}
\nu_{3,4} & =\frac{1}{\sqrt{2}} \sqrt{\mathcal{E}_{11}+\mathcal{E}_{22} \pm \sqrt{(\mathcal{E}_{11}-\mathcal{E}_{22})^{2}+4\mathcal{E}_{12}\mathcal{E}_{21}}}. 
\end{align}
Again, these results are equivalent to the ones obtained by the purification ansatz~\cite{lodewyck2007quantum, fossier2009improvement}.

\subsection{Trusted preparation noise} \label{sec_trustedprep}

\begin{figure}
\centering
\subcaptionbox{}
    [\linewidth]{\includegraphics[width=\linewidth]{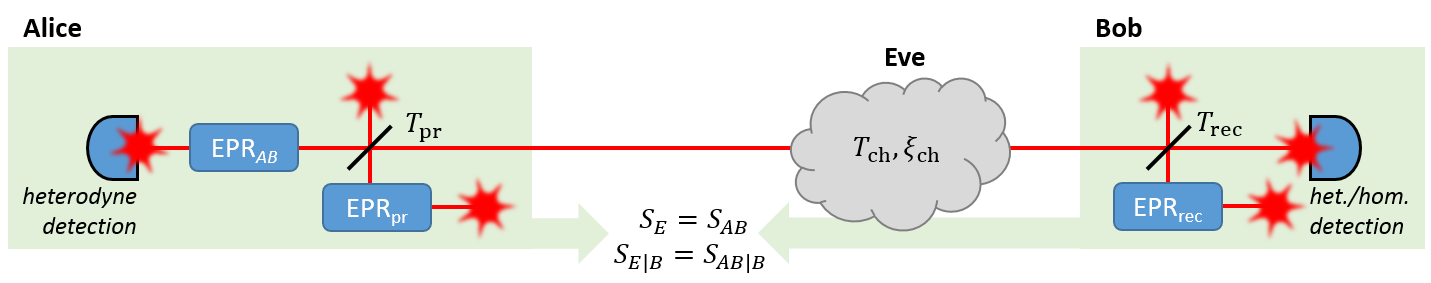}}
\subcaptionbox{}
    [\linewidth]{\includegraphics[width=\linewidth]{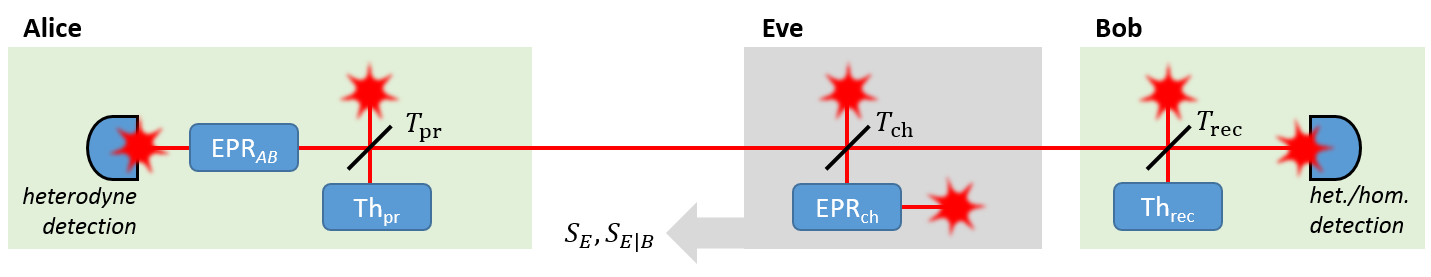}}
\caption{Security analysis under trusted receiver \emph{and} state preparation using (a) the purification and (b) the entangling-cloner ansatz. For the purification ansatz we need to model the preparation noise with an additional entangled state $\text{EPR}_{\text{pr}}$ whereas in the entangling-cloner approach it is sufficient to use a thermal state $\text{Th}_{\text{pr}}$ with variance $W_{\text{pr}}=\xi_{\text{pr}}/(1-T_{\text{pr}})+1$. Although the total state is extended by one additional mode, the mathematical complexity of the security analysis using the entangling-cloner ansatz does not increase significantly since we still only compute the eigenvalues of Eve's subsystem (represented by a $4 \times 4$~matrix which can again be reduced to dimension~$2 \times 2$ using the transformations described above).}
\label{trusted_prep}
\end{figure}

The above derivation of the Holevo bound is particularly handy when not only receiver loss and noise but also preparation noise is considered to be trusted~\cite{usenko2016trusted}. Preparation noise may be composed of laser phase noise and imperfect quadrature modulation~\cite{jouguet2012analyis}. Trusted preparation was first modelled using the purification-based method~\cite{filip2008continuous, usenko2010feasibility} and later by an entangling-cloner~\cite{jacobsen2015continuous}. Similar to the channel and receiver noise, we model the preparation noise using an additional thermal state $\text{Th}_{\text{pr}}$ with variance

\begin{equation}
W_{\text{pr}} =\frac{\xi_{\text{pr}}}{1-T_{\text{pr}}}+1,
\end{equation}
where $\xi_{\text{pr}}$ is the preparation noise in shot-noise units as produced by Alice. The preparation noise as measured by Bob will then be $\xi_{\text{pr},B}=T_{\text{ch}}T_{\text{rec}}\xi_{\text{pr}}$. The thermal state $\text{Th}_{\text{pr}}$ will be interfered with Bob's mode using a beamsplitter of transmission $T_{\text{pr}} \rightarrow 1$ (since imperfect preparation does, in contrast to the channel and receiver, not introduce optical attenuation). Although the limit $T_{\text{pr}} \rightarrow 1$ will lead to $W_{\text{pr}} \rightarrow \infty$, the mode reflected into the channel will be $(1-T_{\text{pr}})W_{\text{pr}}=\xi_{\text{pr}}+1-T_{\text{pr}} \rightarrow \xi_{\text{pr}}$ and is therefore finite and well-defined.

Using the purification ansatz, the entropy $S_{E}$ has to be obtained by the symplectic eigenvalues of a $12 \times 12$ matrix, describing $6$ optical modes: Alice's and Bob's modes and the two modes of $W_{\text{pr}}$ and $W_{\text{rec}}$, respectively (see Figure~\ref{trusted_prep}(a)). Computation of $S_{E|B}$ requires the symplectic eigenvalues of a $10 \times 10$ matrix, describing $5$ optical modes, i.e.\ the total state minus Bob's mode which was measured by heterodyne or homodyne detection. Finding the eigenvalues of this matrix is therefore related to solving a fifth-degree polynomial.

In contrast, using the derivation based on the entangling-cloner attack analogous to the previous section, the problem can again be remodelled to an investigation of merely the two modes accessible to Eve (see Figure~\ref{trusted_prep}(b)), reducing the eigenvalue problem to a second-degree polynomial. This allows us to describe trusted preparation and detection noise by analytical expressions of limited complexity.

The total initial state now includes the EPR state responsible for preparation noise:

\begin{equation}
\Sigma_{\text{tot},0}= \text{EPR}_{AB} \oplus \text{Th}_{\text{pr}} \oplus \text{EPR}_{\text{ch}} \oplus \text{Th}_{\text{rec}}.
\end{equation}
The total beamsplitting operator is now $\text{BS}_{\text{tot}}=\text{BS}_{\text{rec}}\text{BS}_{\text{ch}}\text{BS}_{\text{pr}}$. After the transformation

\begin{align}
\Sigma_{\text{tot}}=\text{BS}_{\text{tot}} \ \Sigma_{\text{tot},0} \ \text{BS}_{\text{tot}}^{T}
\end{align}
we extract the block matrix describing Eve's modes and set $T_{\text{pr}}=1$. This yields

\begin{equation}
\Sigma_{E}=
\left(
\begin{array}{cc}

((1-T_{\text{ch}}) (V+\xi_{\text{pr}}) + T_{\text{ch}}W_{\text{ch}}) \mathbb{1}_{2} &  \sqrt{T_{\text{ch}}} \sqrt{W_{\text{ch}}^2-1} \sigma_{z} \\
 \sqrt{T_{\text{ch}}} \sqrt{W_{\text{ch}}^2-1} \sigma_{z} & W_{\text{ch}} \mathbb{1}_{2} \\
\end{array}
\right),
\end{equation}
which, consistently, is identical to~\eqref{eq_SE} after the substitution $V \rightarrow V + \xi_{\text{pr}}$. The symplectic eigenvalues $\nu_{1}$ and $\nu_{2}$ needed for $S_{E}$ are again obtained by~\eqref{eq_nu12}. In order to compute $\nu_{3}$ and $\nu_{4}$, as required for $S_{E|B}$, we again build the partial covariance matrix of the total state after a projective heterodyne~\eqref{eq_partialhet} or homodyne~\eqref{eq_partialhom} measurement of Bob's mode and then extract the block matrix describing Eve's modes.

After setting $T_{\text{pr}}=1$ Eve's covariance matrix conditioned on Bob's measurement is of the form~\eqref{eq_E|Bhet} in the case of heterodyne detection and of the form~\eqref{eq_E|Bhom} in the case of homodyne detection. Moreover the elements of $\Sigma_{E|B,\text{het}}$ and $\Sigma_{E|B,\text{hom}}$ are identical to \eqref{eq_erechet} (heterodyne) and \eqref{eq_erechom} (homodyne) except for the substitution $V \rightarrow V+\xi_{\text{pr}}$. The symplectic eigenvalues $\nu_{3}$ and $\nu_{4}$ are again obtained by~\eqref{eq_nu34het} (heterodyne), or by~\eqref{eq_nu34hom} (homodyne), respectively. Numerical evaluation of our analytical equations yields a convincing accordance with the numerical results obtained by the purification ansatz~\cite{usenko2016trusted}.

\section{Numerical evaluation} \label{sec_numerical}

\begin{figure}
\centering
\subcaptionbox{}
    [0.495\linewidth]{\includegraphics[width=0.48\linewidth]{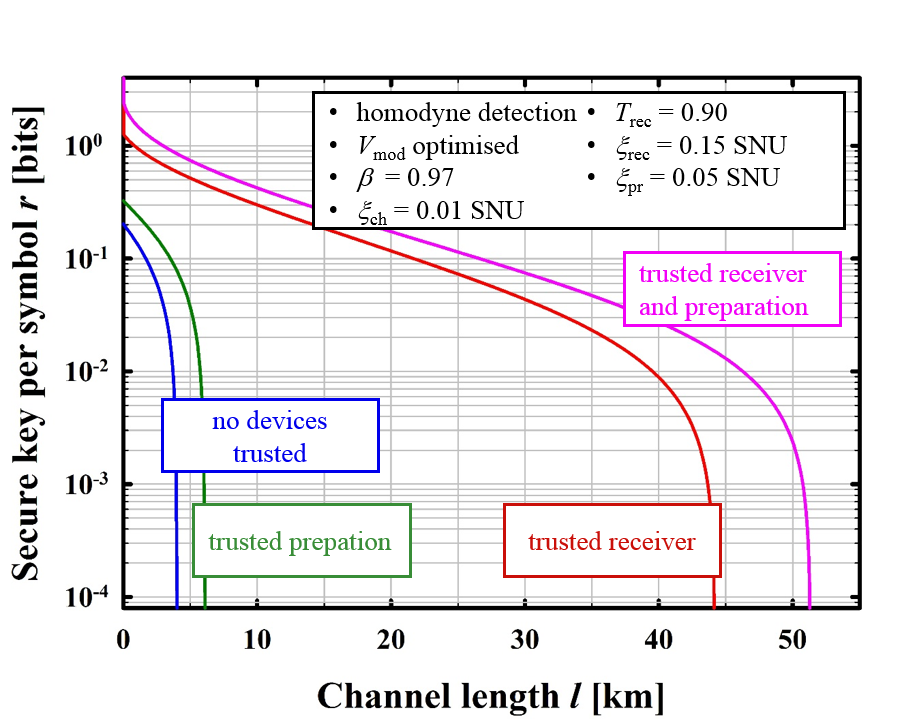}}
\subcaptionbox{}
    [0.495\linewidth]{\includegraphics[width=0.48\linewidth]{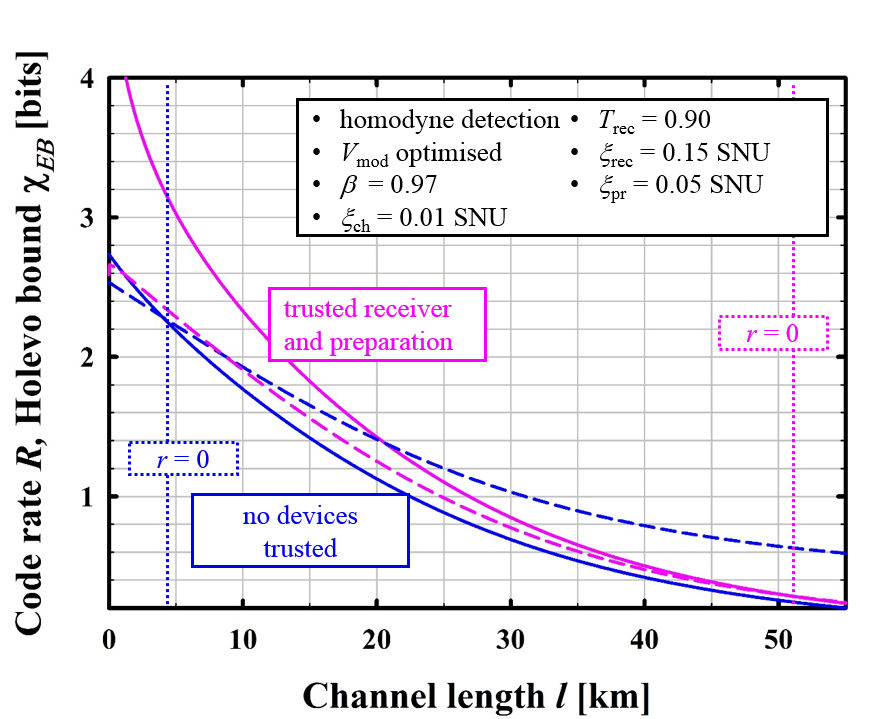}}
\caption{(a) Secure-key rate $r$ with respect to channel length $l$ under different security assumptions for an exemplary set of experimental parameters. The assumed fibre loss is \SI{0.2}{dB/km} and $V_{\text{mod}}:=V-1$ has been optimised to maximise the key rate for each individual point in the graph. (b)~Code rate $R:=\beta I_{AB}$ (solid) and Holevo bound $\chi_{EB}$ (dashed) for the same parameters. The key rate $r$ becomes zero where $R \leq \chi_{EB}$. (In this figure and all subsequent plots we assume zero frame errors and neglect disclosure of samples for parameter estimation.)}
\label{trusted_untrusted}
\end{figure}

\begin{figure}
\centering
\subcaptionbox{}
    [0.495\linewidth]{\includegraphics[width=0.48\linewidth]{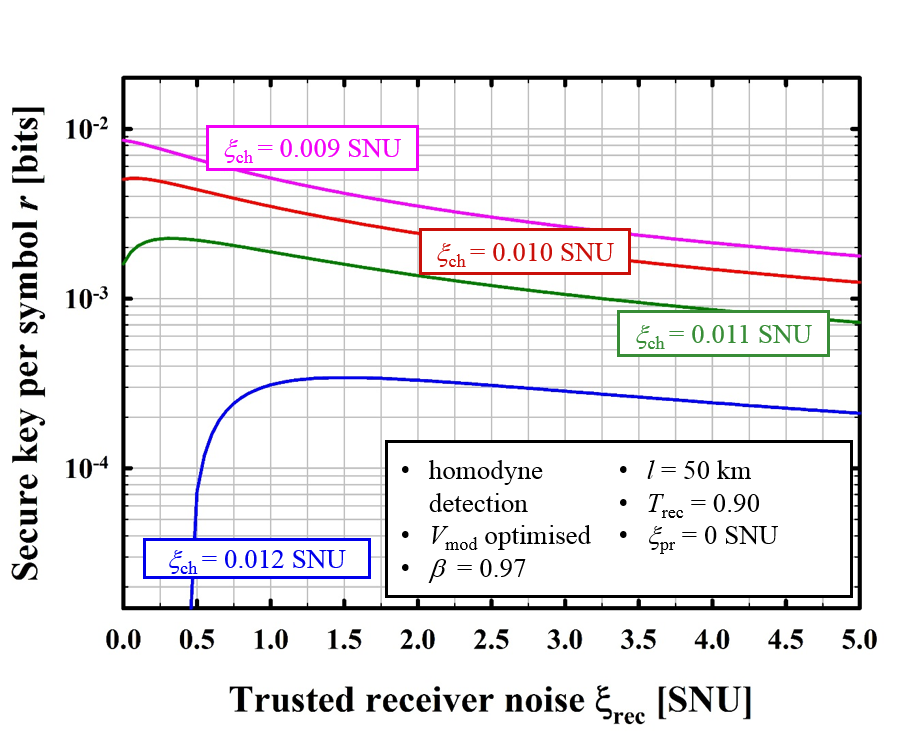}}
\subcaptionbox{}
    [0.495\linewidth]{\includegraphics[width=0.48\linewidth]{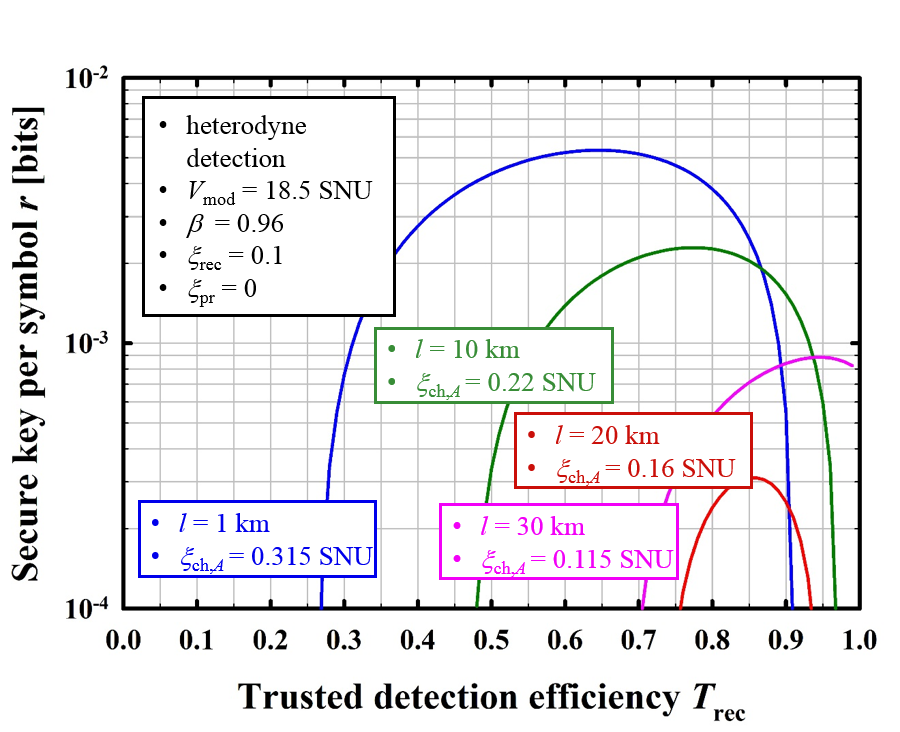}}
\caption{Secure key $r$ with respect to (a) trusted receiver noise $\xi_{\text{rec}}$ and (b) trusted detection efficiency $T_{\text{rec}}$. The graphs are non-monotonous, indicating that under certain conditions a certain amount of noise and detection inefficiency can actually be beneficial for the achievable key rate.}
\label{rec_noise_efficiency}
\end{figure}

\begin{figure}
\centering
\includegraphics[width=0.6\linewidth]{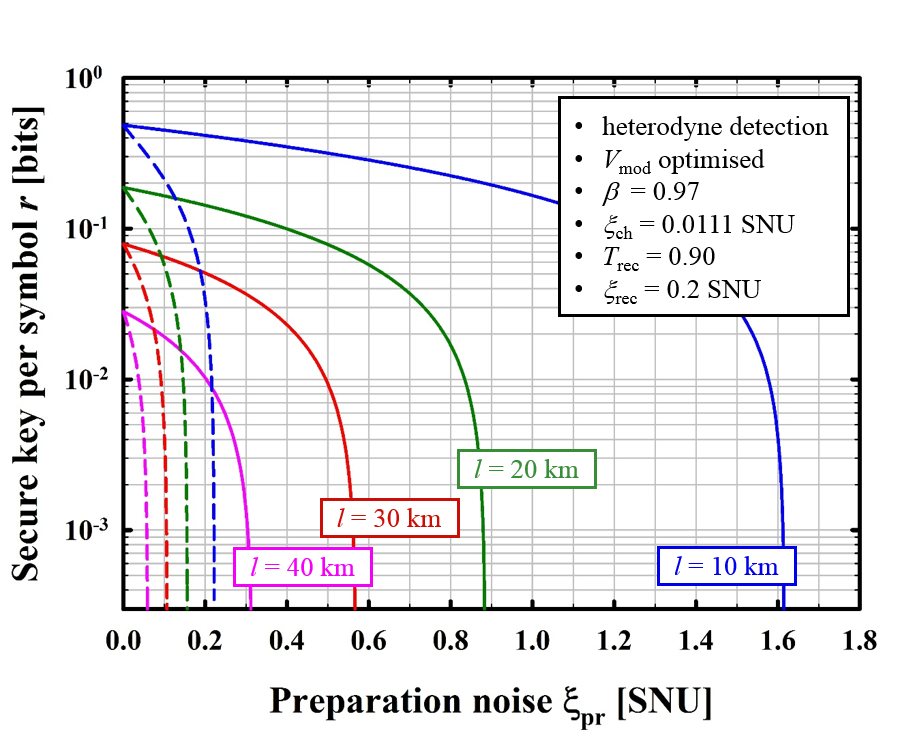}
\caption{Secure key $r$ with respect to trusted (solid) and untrusted (dashed) preparation noise $\xi_{\text{pr}}$, parametrised by channel length $l$. Receiver noise and loss are trusted.}
\label{prep_noise}
\end{figure}

\begin{figure}
\centering
\subcaptionbox{}
    [0.495\linewidth]{\includegraphics[width=0.456\linewidth]{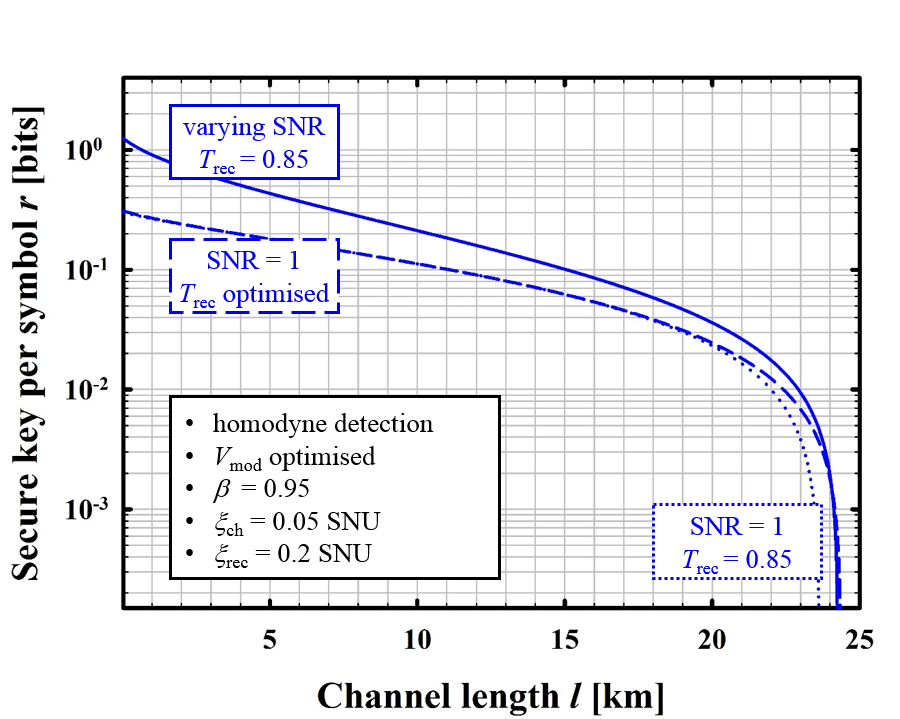}}
\subcaptionbox{}
    [0.495\linewidth]{\includegraphics[width=0.49\linewidth]{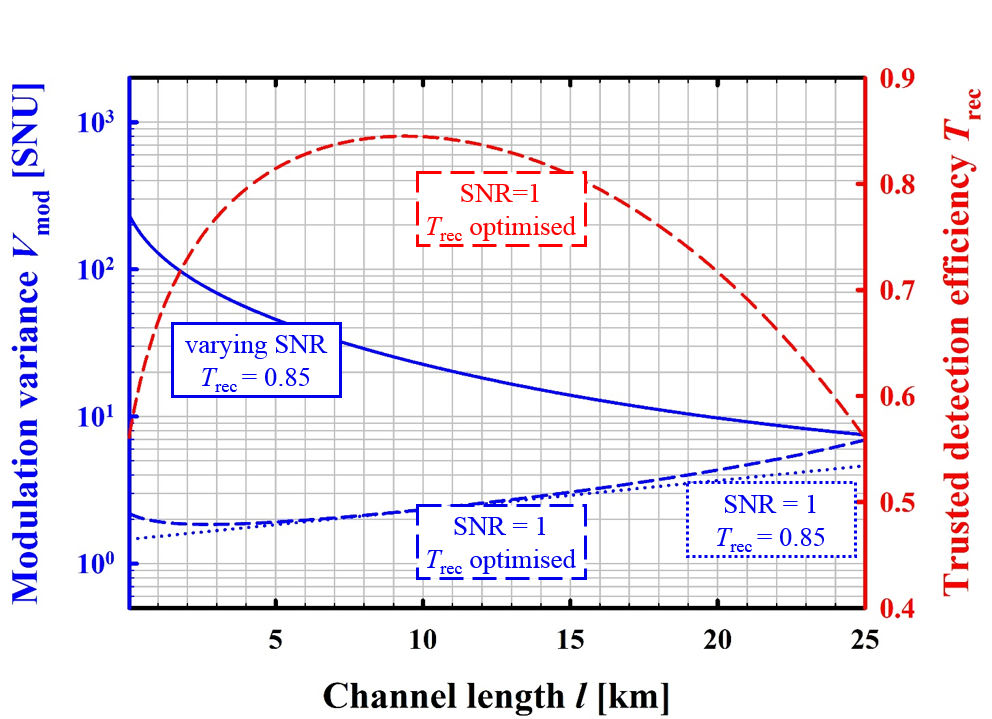}}
\caption{Using trusted receiver loss to keep the SNR constant. This is relevant when one LDPC code (in this example matched to an exemplary SNR of 1) is used for different channel distances. (a)~Comparison of secure key vs.\ channel distance for: key-maximising modulation variance (solid), adjusted $V_{\text{mod}}$ to satisfy $\text{SNR}=1$ (dotted) and optimised $V_{\text{mod}}$ \emph{and} $T_{\text{rec}}$ to maximise the key rate under the condition $\text{SNR}=1$ (dashed). (b)~Modulation variance $V_{\text{mod}}$ for the same three cases under the same experimental parameter set. The red dashed line refers to the right $y$-axis and illustrates the tuning of $T_{\text{rec}}$ to maximise the key while satisfying $\text{SNR}=1$.}
\label{constSNR}
\end{figure}

Using our CV-QKD simulation tool `CVsim'~\cite{laudenbach2016cvsim}, we were able to conduct a thorough study of trusted receiver and state preparation and their implication on experimental implementations. Figure~\ref{trusted_untrusted} indicates the performance difference in terms of the achievable key rate and channel length depending on whether preparation and/or receiver are trusted. Figures~\ref{rec_noise_efficiency}(a) and (b) illustrate the key rate with respect to trusted receiver noise and loss, parametrised by channel noise and channel length. Interestingly, these graphs exhibit a non-monotonous behaviour, indicating that the trusted-receiver assumption does not only render device imperfections less harmful; even more, trusted receiver loss and noise can actually be used to \emph{increase} the key rate~\cite{usenko2016trusted, garcia2009continuous, laudenbach2019noisy}. This is possible when $T_{\text{rec}}$ and $\xi_{\text{rec}}$ decrease the Holevo information $\chi_{EB}$ more than they decrease the mutual information $I_{AB}$. This effect, first observed in DV-QKD~\cite{renner2005information} and sometimes described by `fighting noise with noise'~\cite{garcia2007quantum} is only possible in the trusted-receiver scenario where imperfect detection increases the conditional entropy $S_{E|B}$ but leaves the entropy $S_{E}$ invariant. Figure~\ref{prep_noise} illustrates the secure-key rate with respect to trusted and untrusted preparation noise for different channel lengths.

As a further consideration, deliberately detuning the receiver's trusted quantum efficiency $T_{\text{rec}}$ can become relevant in the practical case where the error-correcting code (usually low-density parity-check, LDPC) is optimised to a particular signal-to-noise ratio. If one and the same LDPC code is supposed to be used for different channel distances then Alice and Bob will need to keep the SNR constant with respect to channel loss. This can be achieved in a straightforward manner by tuning the modulation variance $V_{\text{mod}}$ accordingly, or alternatively, by adjusting $V_{\text{mod}}$ \emph{and} $T_{\text{rec}}$. Figure~\ref{constSNR}(a) compares the key rate over channel length for an exemplary parameter set in three cases: (1)~modulation variance optimised to maximise the secure key, irregardless of the SNR (solid), (2)~modulation variance adjusted to keep the SNR constant at 1 (dotted) and (3)~modulation variance \emph{and} receiver efficiency optimised to maximise the key rate and, at the same time, satisfy $\text{SNR}=1$ (dashed). We observe (for this particular parameter set) a slight improvement in terms of accessible channel length when $T_{\text{rec}}$ is deliberately detuned. Figure~\ref{constSNR}(b) compares the modulation variance for each of these cases and illustrates the optimal quantum efficiency $T_{\text{rec}}$ under the condition of a constant SNR.

Certainly enough, the more common procedure is to adapt the LDPC code for forward error correction to a certain SNR level corresponding to the link. In order to match the code rate accurately to the SNR in the presence of fluctuations, techniques like puncturing and shortening are usually used~\cite{richardson2008modern, elkouss2015rate}. Our proposal to deliberately reduce the detection efficiency can be seen as an alternative way of code-rate matching. We emphasise the observation that varying $V_{\text{mod}}$ and $T_{\text{rec}}$ (Fig.~\ref{constSNR}(a), dashed blue) allows us to implement one LDPC code, matched to a constant SNR, that can produce a secure key with no distance penalty compared to codes that are matched to the `natural’ SNR corresponding to the respective channel length (Fig.~\ref{constSNR}, solid blue). This is in contrast to adapting the code itself which comes at the cost of a reduced reconciliation efficiency $\beta$. Especially at long distances even a slight decrease of $\beta$ can reduce the secret fraction $r \sim \beta I_{AB}-\chi_{EB}$ to the negative regime.

From an application point of view, adjusting the SNR by increasing the trusted detector loss or, alternatively, the trusted receiver noise~\cite{kreinberg2019adding} can become relevant when experimental requirements demand for more flexibility in terms of the SNR. For instance, these may include the implementation of different security assumptions (and correspondingly adapted signal powers) at one and the same link, dynamic switching between homodyne and heterodyne detection, field tests and multi-node networks with different distances.

\section{Conclusion} \label{sec_concl}

In conclusion, we demonstrated and derived in detail an efficient way to compute the Holevo bound in continuous-variable quantum key distribution under the assumption of trusted receiver and state preparation. In particular, we showed that the eigenvalue problem that needs to be solved in order to obtain the Holevo bound can be reduced to a second-degree polynomial, regardless of the total system's complexity. This allowed us to find analytical expressions for Eve's entropy and conditional entropy under the assumption of trusted receiver \emph{and} state preparation. Finally, we performed numerical simulations to illustrate the impact of various trusted-device assumptions on practical CV-QKD implementations, highlighting the fact that under particular circumstances the key rate can even be increased by an appropriate amount of trusted receiver loss and noise.

\section*{Acknowledgements}

We thank V.~Usenko, A.~Poppe, K.~Jaksch and K.~Günther for valuable discussions.
 
\vspace{4pt}
\begin{minipage}{0.068\linewidth}
\includegraphics[scale=0.05]{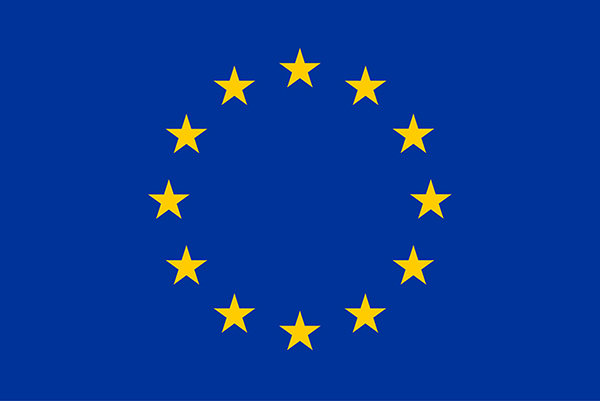}
\end{minipage} 
\begin{minipage}{0.931\linewidth}
This work has received funding from the European Union’s Horizon 2020 research and innovation programme through the Quantum-Flagship projects UNIQORN (no. 820474) and CiViQ (no. 820466).
\end{minipage}

%% file: trustednoise_arxiv_v3.bbl
\begin{thebibliography}{99}

\bibitem{laudenbach2017continuous} F. Laudenbach, C. Pacher, C.-H. F. Fung, A. Poppe, M. Peev, B. Schrenk, M. Hentschel, P. Walther, and H. H\"ubel, \textit{Continuous-Variable Quantum Key Distribution with Gaussian Modulation -- The Theory of Practical Implementations}, Adv. Quantum Technol. \textbf{1}, 1800011 (2018).

\bibitem{grosshans2002continuos} F.  Grosshans  and  P.  Grangier,  \textit{Continuous  variable  quantum  cryptography  using  coherent  states}, Phys. Rev. Lett. \textbf{88}, 057902 (2002).

\bibitem{grosshans2003quantum} F. Grosshans, G. V. Assche, J. Wenger, R. Brouri, N. J. Cerf, and P. Grangier, \textit{Quantum key distribution using Gaussian-modulated coherent states}, Nature \textbf{421}, 238 (2003).

\bibitem{grosshans2003virtual} F. Grosshans, N. J. Cerf, J. Wenger, R. Tualle-Brouri, and  P.  Grangier, \textit{Virtual  entanglement and reconciliation protocols for quantum cryptography with continuous variables}, Quantum Inf. Comput. \textbf{3}, 535 (2003).

\bibitem{weedbrook2004quantum} C. Weedbrook, A. M. Lance, W. P. Bowen, T. Symul, T. C. Ralph, and P. K. Lam, \textit{Quantum cryptography without switching}, Phys. Rev. Lett. \textbf{93}, 170504 (2004).

\bibitem{cerf2001quantum} N. J. Cerf, M. Lévy, and G. Van Assche, \textit{Quantum distribution of Gaussian keys using squeezed states}, Phys. Rev. A \textbf{63}, 052311 (2001).

\bibitem{madsen2012continuos} L. S. Madsen, V. C. Usenko, M. Lassen, R. Filip, and U. L. Andersen, \textit{Continuous variable quantum key distribution with modulated entangled states}, Nat. Comm. \textbf{3}, 1083 (2012).

\bibitem{devetak2004distillation} I. Devetak and A. Winter, \textit{Distillation of secret key and entanglement from quantum states}, P. Roy. Soc. Lond. A Mat. \textbf{461}, 207 (2004).

\bibitem{weedbrook2012gaussian} C. Weedbrook, S. Pirandola, R. García-Patrón, N. J. Cerf, T. C. Ralph, J. H. Shapiro, and S. Lloyd, \textit{Gaussian quantum information}, Rev. Mod. Phys. \textbf{84}, 621 (2012).

\bibitem{nielsen2000quantum} M. A. Nielsen and I. L. Chuang, \textit{Quantum Computation and Quantum Information}, Cambridge University Press, Cambridge, GBR (2000).

\bibitem{lodewyck2007quantum} J.~Lodewyck, M.~Bloch, R.~Garc\'{\i}a-Patr\'on, S.~Fossier, E.~Karpov, E.~Diamanti, T.~Debuisschert, N.~J. Cerf, R.~Tualle-Brouri, S.~W. McLaughlin, and P.~Grangier, \textit{Quantum key distribution over 25 km with an all-fiber continuous-variable system}, Phys. Rev. A \textbf{76}, 042305 (2007).

\bibitem{fossier2009improvement} S. Fossier, E. Diamanti, T. Debuisschert, R. Tualle-Brouri, and P. Grangier, \textit{Improvement of continuous-variable quantum key distribution systems by using optical preamplifiers}, J. Phys. B \textbf{42}, 114014 (2009).

\bibitem{usenko2016trusted} V.~C. Usenko and R.~Filip, \textit{Trusted noise in continuous-variable quantum key distribution: A threat and a defense}, Entropy \textbf{18}, 20 (2016).

\bibitem{weedbrook2012continuous} C. Weedbrook, S. Pirandola, and T. C. Ralph, \textit{Continuous-variable quantum key distribution using thermal states}, Phys. Rev. A \textbf{86}, 022318 (2012).

\bibitem{jouguet2012analyis} P. Jouguet, S. Kunz-Jacques, E. Diamanti, and A. Leverrier, \textit{Analysis of imperfections in practical continuous-variable quantum key distribution}, Phys. Rev. A \textbf{86}, 032309 (2012).

\bibitem{filip2008continuous} R. Filip, \textit{Continuous-variable quantum key distribution with noisy coherent states}, Phys. Rev. A \textbf{77}, 022310 (2008).

\bibitem{usenko2010feasibility} V.~C. Usenko and R. Filip, \textit{Feasibility of continuous-variable quantum key distribution with noisy coherent states}, Phys. Rev. A \textbf{81}, 022318 (2010).

\bibitem{jacobsen2015continuous} C. S. Jacobsen, T. Gehring, and U. L. Andersen, \textit{Continuous Variable Quantum Key Distribution with a Noisy Laser}, Entropy \textbf{17}, 4654 (2015).

\bibitem{laudenbach2016cvsim} F.~Laudenbach, C.~Pacher, C.-H.~F. Fung, M.~Peev, A.~Poppe, and H.~H{\"u}bel, \textit{CVsim -- a novel CVQKD simulation tool}, Proc.\ 6th International Conference on Quantum Cryptography (Washington DC, USA, 2016), Poster176.

\bibitem{garcia2009continuous} R. García-Patrón and N.~J. Cerf, \textit{Continuous-Variable Quantum Key Distribution Protocols Over Noisy Channels}, Phys. Rev. Lett. \textbf{102}, 130501 (2009).

\bibitem{laudenbach2019noisy} F. Laudenbach and C. Pacher, \textit{Noisy Detector? Good! Analysis of Trusted-Receiver Scenario in Continuous-Variable Quantum Key Distribution}, Proc. Quantum Information and Measurement V (Rome, ITA, 2019), T5A.59.

\bibitem{renner2005information} R. Renner, N. Gisin, and B. Kraus, \textit{Information-theoretic security proof for quantum-key-distribution protocols}, Phys. Rev. A \textbf{72}, 012332 (2005).

\bibitem{garcia2007quantum} R. García-Patrón, \textit{Quantum information with optical continuous variables: from Bell tests to key distribution}, PhD thesis, Université libre de Bruxelles, 2007.

\bibitem{richardson2008modern} T. Richardson and R. Urbanke, \textit{Modern Coding Theory}, Cambridge University Press, Cambridge, GBR (2008).

\bibitem{elkouss2015rate} D. Elkouss, J. Martinez, D. Lancho, and V. Martin, \textit{Rate compatible protocol for information reconciliation: An application to QKD}, Proc. IEEE Information Theory Workshop (Cairo, EGY, 2010), 1.

\bibitem{kreinberg2019adding} S. Kreinberg, I. Koltchanov, and A. Richter, \textit{Adding artificial noise for code rate matching in continuous-variable quantum key distribution}, arxiv:1905.04925 (2019).

\end{thebibliography}
